\documentclass[12pt,a4j]{article} 
\makeatletter
\newdimen{\eqarcolsep}
\usepackage[dvips]{color,graphicx} 
\usepackage{amsmath} 
\usepackage{amsthm} 
\usepackage{bm} 
\usepackage{fancybox} 
\usepackage{ascmac} 
 \usepackage{amssymb}

\makeatletter 
\begin{document}

{\large \bf Regular biorthogonal pairs and Psuedo-bosonic operators}

\ \\
\begin{center}
Hiroshi Inoue and Mayumi Takakura\\
\end{center}

\ \\
\begin{abstract} 
The first purpose of this paper is to show a method of constructing a regular biorthogonal pair based on the commutation rule: $ab-ba=I$ for a pair of  operators $a$ and $b$ acting on a Hilbert space ${\cal H}$ with inner product $( \cdot | \cdot )$. Here, sequences $\{ \phi_{n} \}$ and $\{ \psi_{n} \}$ in a Hilbert space ${\cal H}$ are biorthogonal if $( \phi_{n} | \psi_{m})= \delta_{nm}$, $n,m=0,1, \cdots$, and they are regular if both $D_{\phi} \equiv Span \{ \phi_{n} \}$ and $D_{\psi} \equiv Span \{ \psi_{n} \}$ are dense in ${\cal H}$. Indeed, the assumption to construct the regular biorthogonal pair coincide with the definition of pseudo-bosons as originally given in Ref \cite{bagarello10}. Furthermore, we study the connections between the pseudo-bosonic operators $a, \; b, \; a^{\dagger}, \; b^{\dagger}$ and the pseudo-bosonic operators defined by a regular biorthogonal pair $(\{ \phi_{n} \}$, $\{ \psi_{n} \} )$ and an ONB $\bm{e}$ of ${\cal H}$ in appeared Ref \cite{hiroshi1}. The second purpose is to define and study the notion of ${\cal D}$-pseudo bosons in Ref \cite{bagarello13, bagarello2013} and give a method of constructing ${\cal D}$-pseudo bosons on some steps. Then it is shown that for any ONB $\bm{e}= \{ e_{n} \}$ in ${\cal H}$ and any operators $T$ and $T^{-1}$ in ${\cal L}^{\dagger} ( {\cal D})$, we may construct operators $A$ and $B$ satisfying ${\cal D}$-pseudo bosons, where ${\cal D}$ is a dense subspace in a Hilbert space ${\cal H}$ and ${\cal L}^{\dagger} ( {\cal D})$ the set of all linear operators $T$ from ${\cal D}$ to ${\cal D}$ such that $T^{\ast} {\cal D} \subset {\cal D}$, where $T^{\ast}$ is the adjoint of $T$. Finally, we give some physical examples of ${\cal D}$-pseudo bosons based on standard bosons by the method of constructing ${\cal D}$-pseudo bosons stated above.\\
\end{abstract}

\section{Introduction}
In this paper, we shall show a method of constructing a regular biorthogonal pair based on the following commutation rule under some assumptions. Here, the commutation rule is that a pair of operators $a$ and $b$ acting on a Hilbert space ${\cal H}$ with inner product $( \cdot | \cdot )$ satisfies $ab-ba=I$. Indeed, the assumptions to construct the regular biorthogonal pair coincide with the definition of pseudo-bosons as originally given in Ref \cite{bagarello10}, where in the recent literature many researchers have investigated in Ref \cite{bagarello13, bagarello10, bagarello11, mostafazadeh, d-t}. Furthermore, in Ref \cite{hiroshi1} the author have studied the general theory of operators $A_{\bm{e}}, B_{\bm{e}},A_{\bm{e}}^{\dagger},B_{\bm{e}}^{\dagger}, N_{\bm{e}},N_{\bm{e}}^{\dagger}$ defined by a regular biorthogonal pair $(\{ \phi_{n} \}$, $\{ \psi_{n} \} )$ and an ONB $\bm{e}$ in ${\cal H}$. Though we will describe the formulas of these operators $A_{\bm{e}}, B_{\bm{e}},A_{\bm{e}}^{\dagger},B_{\bm{e}}^{\dagger}, N_{\bm{e}},N_{\bm{e}}^{\dagger}$ in detail in Section 2, the operators connect with ${\it quasi}$-${\it hermitian \; quantum \; mechanics}$ and its relatives. Many researchers have investigated such operators mathematically in Ref \cite{h-t, b-i-t, hiroshi1} Therefore, we study the connections between the operators $a, \; b, \; a^{\dagger}, \; b^{\dagger}$ and the operators  $A_{\bm{e}}$, $B_{\bm{e}}$, $A_{\bm{e}}^{\dagger}, \; B_{\bm{e}}^{\dagger}$. By Proposition 2.1, (3) and Proposition 2.2, (3), we show that $\{ a,b \}$ and $\{ a^{\dagger}, b^{\dagger} \}$ have an algebraic structure, respectively, that is, algebras generated by $\{ 1, a \lceil_{D_{\phi}}, b\lceil_{D_{\phi}} \}$ and by $\{ 1, a^{\dagger} \lceil_{D_{\psi}}, b^{\dagger} \lceil_{D_{\psi}} \}$ are defined, respectively. However, $\{ 1,a,b,a^{\dagger},b^{\dagger} \}$ do not have a $
\ast$-algebraic structure, in general. From this reason, the second purpose is to define and study the notion of ${\cal D}$-pseudo bosons.\cite{bagarello13, bagarello2013} Here, ${\cal D}$ is a dense subspace in a Hilbert space ${\cal H}$. We denote by ${\cal L}^{\dagger} ( {\cal D})$ the set of all linear operators $T$ from ${\cal D}$ to ${\cal D}$ such that $T^{\ast} {\cal D} \subset {\cal D}$, where $T^{\ast}$ is the adjoint of $T$. ${\cal L}^{\dagger} ( {\cal D})$ is a $\ast$-algebra under the usual operators $\alpha T$, $S+T$, $ST$ and an involution $T^{\dagger} \equiv T^{\ast} \lceil_{{\cal D}}$, and it is called a maximal ${\cal O}^{\ast}$-algebra on ${\cal D}$.\cite{k-s} A pair of operators $a$ and $b$ is said ${\cal D}$-pseudo bosons if $a$, $b \in {\cal L}^{\dagger}( {\cal D})$ satisfy following conditions (i), (ii) and (iii): \\
\hspace{3mm} (i) $ab-ba =I$. \\
\hspace{3mm} (ii) There exists a non-zero element $\phi_{0} \in {\cal D}$ such that $a\phi_{0}=0$. \\
\hspace{3mm} (iii) There exists a non-zero element $\psi_{0} \in {\cal D}$ such that $b^{\dagger} \psi_{0}=0$.\\
Furthermore, we show a method of constructing ${\cal D}$-pseudo bosons on some steps. Then it is shown that for any ONB $\bm{e}= \{ e_{n} \}$ in ${\cal H}$ and any operators $T$ and $T^{-1}$ in ${\cal L}^{\dagger} ( {\cal D})$, we may construct operators $A$ and $B$ satisfying ${\cal D}$-pseudo bosons. Finally, we give some physical examples of ${\cal D}$-pseudo bosons based on a method of constructing ${\cal D}$-pseudo bosons for some steps.

This article is organized as follows. In Section 2, we introduce a method of constructing a regular biorthogonal pair based on the commutation rule under some assumption. And we investigate the connections between the operators $a, \; b, \; a^{\dagger}, \; b^{\dagger}$ and the operators  $A_{\bm{e}}$, $B_{\bm{e}}$, $A_{\bm{e}}^{\dagger}, \; B_{\bm{e}}^{\dagger}$. In Section 3, we define and study the notion of ${\cal D}$-pseudo bosons and a method of constructing ${\cal D}$-pseudo bosons for some steps. In Section 4, we give some physical examples of ${\cal D}$-pseudo bosons based on the method of constructing ${\cal D}$-pseudo bosons in Section 3.
\section{Regular biorthogonal pairs and Pseudo-bosons}
Let ${\cal H}$ be a Hilbert space with inner product $( \cdot | \cdot )$. We introduce a pair of operators $a$ and $b$ acting on ${\cal H}$ satisfying the following commutation rules
\begin{eqnarray}
ab-b a=I . \nonumber
\end{eqnarray}
In particular, this collapses to the canonical commutation rule (CCR) if $b= a^{\dagger}$. Furthermore, we introduce the notions of biorthogonal sequences and the regularity as follows: Sequences $\{ \phi_{n} \}$ and $\{ \psi_{n} \}$ in a Hilbert space ${\cal H}$ are biorthogonal if $( \phi_{n} | \psi_{m})= \delta_{nm}$, $n,m=0,1, \cdots$, and they are regular if both $Span \{ \phi_{n} \}$ and $Span \{ \psi_{n} \}$ are dense in ${\cal H}$. Then $( \{ \phi_{n} \} , \{ \psi_{n} \} )$
 is said to be a regular biorthogonal pair. We construct biorthogonal pairs based on the above commutation rule under some assumptions. At first, we assume the following statement.\\
\par
{\it Assumption 1.} {\it There exists a non-zero element $\phi_{0}$ of ${\cal H}$ such that \\
\hspace{3mm} (i) $a \phi_{0}=0$, \\
\hspace{3mm} (ii) $\phi_{0} \in D^{\infty}(b) \equiv \cap_{k=0}^{\infty} D( b^{k})$, \\
\hspace{3mm} (iii) $b^{n} \phi_{0} \in D(a)$, $n=0,1, \cdots$.}\\\\
Then, we may define a sequence $\{ \phi_{n} \}$ in ${\cal H}$ by
\begin{eqnarray}
\phi_{n} 
&\equiv& \frac{1}{\sqrt{n!}} \; b^{n} \phi_{0}, \;\;\; n =0,1, \cdots \nonumber \\
&=& \frac{1}{\sqrt{n}} \; b \phi_{n-1}, \;\;\; n =1,2, \cdots . \nonumber 
\end{eqnarray}
If we only define the above sequence $\{ \phi_{n} \}$ in ${\cal H}$, we do not need Assumption 1, (iii). However, we need Assumption 1, (iii) to define the operator $N \equiv b a$. Then we have the following\\
\par
{\it Proposition 2.1.} {\it The following statements hold. \\
\hspace{3mm} (1) $b^{n} \phi_{0} \in D(a^{m})$ and 
\begin{eqnarray}
a^{m} b^{n} \phi_{0}
&=& \left\{
\begin{array}{cl}
_{n}P_{m} b^{n-m} \phi_{0} \;\;\;\;\;\;\;\;\;\; &,m\leq n, \\
\nonumber \\
0 \;\;\; &,m > n.
\end{array}
\right. \nonumber 
\end{eqnarray}
\hspace{3mm} (2) $\phi_{n} \in D(N^{m})$ and $N^{m} \phi_{n}=n^{m} \phi_{n}$, $n,m=0,1, \cdots$. In particular, $N\phi_{n}=n \phi_{n}$, $n=0,1, \cdots$.\\
\hspace{3mm} (3) 
\begin{eqnarray}
a \phi_{n}
&=& \left\{
\begin{array}{cl}
0 \;\;\;\;\;\;\;\;\;\; &,n=0, \\
\nonumber \\
\sqrt{n} \phi_{n-1}, \;\;\; &,n=1,2, \cdots,
\end{array}
\right. \nonumber \\
\nonumber \\
b \phi_{n} &=& \sqrt{n+1} \phi_{n+1}  \;\;\;\;\;\;\;\;\;\;\; ,n=0,1, \cdots .\nonumber 
\end{eqnarray} }
\par
{\it Proof.} (1) We prove this statement based on mathematical induction. We prove this statement when $m=1$:
\begin{equation}
ab^{n} \phi_{0}
= \left\{
\begin{array}{cl}
nb^{n-1}\phi_{0} \;\;\; &,n=1,2, \cdots , \\
\nonumber \\
0 \;\;\;\;\;\;\;\;\;\;\; &,n=0. \\
\end{array}
\right. \tag{2.1}
\end{equation}
Since $a\phi_{0}=0$ by Assumption 1, (i), (2.1) holds when $n=0$. Let $n=1$. Since $\phi_{0} \in D(ab) \cap D(ba)$ by Assumption 1, (i), (iii), we have
\begin{equation}
ab\phi_{0}= (ba+1)\phi_{0}=\phi_{0}. \tag{2.2}
\end{equation}
Hence (2.1) holds when $n=1$. Assume that (2.1) holds when $n=k \geq 2$, that is,
\begin{equation}
b^{k}\phi_{0} \in D(a) \;\;\; {\rm and} \;\;\; ab^{k}\phi_{0}=kb^{k-1} \phi_{0}. \tag{2.3}
\end{equation}
Let $n=k+1$. Then we have $b^{k}\phi_{0} \in D(ab) \cap D(ba)$ by Assumption 1, (iii) and (2.3), which implies by making use of the commutation relation $ab-ba=I$ that
\begin{eqnarray}
ab^{k+1} \phi_{0}
&=& ab ( b^{k}\phi_{0}) \nonumber \\
&=& (ba+I) b^{k}\phi_{0} \nonumber \\
&=& bab^{k} \phi_{0}+ b^{k} \phi_{0} \nonumber \\
&=& (k+1) b^{k} \phi_{0}. \nonumber 
\end{eqnarray}
Thus the statement (2.1) holds. Assume that when $m=k$ the statement (2.1) holds: $b^{n}\phi_{0} \in D(a^{k})$ and
\begin{equation}
a^{k}b^{n} \phi_{0}
= \left\{
\begin{array}{cl}
_{n}P_{k} b^{n-k}\phi_{0} \;\;\; &,k \leq n , \\
\nonumber \\
0 \;\;\;\;\;\;\;\;\;\;\; &,k>n. \\
\end{array}
\right. \tag{2.4}
\end{equation}
We show the statement (2.1) when $m=k+1$. Let $n \geq k+1$. By (2.4), we have $b^{n}\phi_{0} \in D(a^{k+1})$ and furthermore, by (2.4) and (2.1)
\begin{eqnarray}
a^{k+1}b^{n}\phi_{0}
&=& a (a^{k}b^{n}\phi_{0}) \nonumber \\
&=& a( \; _{n}P_{k} b^{n-k} \phi_{0} ) \nonumber \\
&=& \; _{n}P_{k} (n-k) b^{n-k-1} \phi_{0} \nonumber \\
&=& \; _{n}P_{k+1} b^{n-(k+1)}\phi_{0}. \nonumber
\end{eqnarray}
Let $n=k$. By (2.4) we have $b^{k}\phi_{0} \in D(a^{k+1})$ and 
\begin{eqnarray}
a^{k+1}b^{k}\phi_{0}
&=& a ( a^{k}b^{k} \phi_{0}) \nonumber \\
&=& k! a\phi_{0} =0 . \nonumber 
\end{eqnarray}
Let $n<k$. By (2.4) we have
\begin{eqnarray}
a^{k+1}b^{n}\phi_{0}=a (a^{k}b^{n}\phi_{0})=0. \nonumber
\end{eqnarray}
Thus the statement (2.1) holds when $m=k+1$. This completes the proof of (1).\\\\
(2) We prove this statement in case of $m=1$. It follows from (1) that $\phi_{n} \in D(N)$ and
\begin{eqnarray}
N\phi_{n}
&=& ba \left( \frac{1}{\sqrt{n!}} b^{n} \phi_{0 } \right) \nonumber \\
&=& b \left( \frac{1}{\sqrt{n !}} nb^{n-1} \phi_{0} \right) \nonumber \\
&=& n\phi_{n}. \nonumber
\end{eqnarray}
Thus by the above statement and (1), it is  easily shown that $\phi_{n} \in D(N^{m})$ and $N^{m} \phi_{n}=n^{m} \phi_{n}$, $n,m=0,1, \cdots$. \\\\
(3) It follows from (1) and (2). This completes the proof.\\\\\\
The statement of Proposition 2.1, (2) means that $N$ is the number operator for $\{ \phi_{n} \}$. Next we assume the following statement.\\
\par
{\it Assumption 2.} {\it There exists a non-zero element $\psi_{0}$ of ${\cal H}$ such that\\
\hspace{3mm} (i) $b^{\dagger} \psi_{0}=0$, \\
\hspace{3mm} (ii) $\psi_{0} \in D^{\infty}(a^{\dagger}) \equiv \cap_{k=0}^{\infty} D( (a^{\dagger})^{k})$, \\
\hspace{3mm} (iii) $(a^{\dagger})^{n} \psi_{0} \in D(b^{\dagger})$, $n=0,1, \cdots$.}\\\\
Then, we may define a sequence $\{ \psi_{n} \}$ in ${\cal H}$ by
\begin{eqnarray}
\psi_{n} 
&\equiv& \frac{1}{\sqrt{n!}} \; (a^{\dagger})^{n} \psi_{0}, \;\;\; n= 0,1, \cdots \nonumber \\
&=& \frac{1}{\sqrt{n}} \; a^{\dagger} \psi_{n-1}, \;\;\; n =1,2, \cdots . \nonumber 
\end{eqnarray}
And we put an operator $N^{\dagger} \equiv a^{\dagger} b^{\dagger}$. Then we have the following\\
\par
{\it Proposition 2.2.} {\it The following statements hold. \\
\hspace{3mm} (1) $(a^{\dagger})^{n} \psi_{0} \in D((b^{\dagger})^{m})$ and 
\begin{eqnarray}
(b^{\dagger})^{m} (a^{\dagger})^{n} \psi_{0}
&=& \left\{
\begin{array}{cl}
_{n}P_{m} (a^{\dagger})^{n-m} \psi_{0} \;\;\;\;\;\;\;\;\;\; &,m\leq n, \\
\nonumber \\
0 \;\;\; &,m > n.
\end{array}
\right. \nonumber 
\end{eqnarray}
\hspace{3mm} (2) $\psi_{n} \in D((N^{\dagger})^{m})$ and $(N^{\dagger})^{m} \psi_{n}=n^{m} \psi_{n}$, $n,m=0,1, \cdots$. In particular, $N^{\dagger} \psi_{n}=n \psi_{n}$, $n=0,1, \cdots$.\\
\hspace{3mm} (3) 
\begin{eqnarray}
a^{\dagger} \psi_{n} &=& \sqrt{n+1} \psi_{n+1}  \;\;\;\;\;\;\;\;\;\;\; ,n=0,1, \cdots ,\nonumber \\
\nonumber \\
b^{\dagger} \psi_{n}
&=& \left\{
\begin{array}{cl}
0 \;\;\;\;\;\;\;\;\;\; &,n=0, \\
\nonumber \\
\sqrt{n} \psi_{n-1}, \;\;\; &,n=1,2, \cdots .
\end{array}
\right. \nonumber  
\end{eqnarray}}\\
\par
{\it Proof.} This is shown similarly to Proposition 2.1.\\\\
Here, we consider about the sequences $\{ \phi_{n} \}$ and $\{ \psi_{n} \}$. It is easily shown that $( \phi_{n} | \psi_{m})= ( \phi_{0} | \psi_{0}) \delta_{nm}$. From now on, we may assume that $( \phi_{0} | \psi_{0})=1$ without loss of generality. Then $\{ \phi_{n} \}$ and $\{ \psi_{n} \}$ are biorthogonal. Furthermore, we assume the following statements with respect to $\{ \phi_{n} \}$ and $\{ \psi_{n} \}$.\\
\par
{\it Assumption 3.} {\it \\
\hspace{3mm} (i) $Span \{ b^{n} \phi_{0} \}$ is dense in ${\cal H}$.\\
\hspace{3mm} (ii) $Span \{ (a^{\dagger})^{n} \psi_{0} \}$ is dense in ${\cal H}$.}\\\\
Then $D_{\phi} \equiv Span \{ \phi_{n} \} = Span \{ b^{n} \phi_{0} \}$ and $D_{\psi} \equiv Span \{ \psi_{n} \} =Span \{ (a^{\dagger})^{n} \psi_{0} \}$ are dense in ${\cal H}$.  If a pair of operator $a$ and $b$ acting on ${\cal H}$ satisfy Assumption 1-3, $( \{ \phi_{n} \} , \{ \psi_{n} \} )$ becomes a regular biorthogonal pair. The assumption to construct the regular biorthogonal pair coincide with the definition of pseudo-bosons as originally given in Ref \cite{bagarello10}.

In Ref \cite{hiroshi1}, the author have studied the general theory of regular biorthogonal pairs. In particular, it has been shown that if $( \{ \phi_{n} \} , \{ \psi_{n} \} )$ is a regular biorthogonal pair in a Hilbert space ${\cal H}$, then for any ONB $\bm{e} = \{e_{n} \}$ in ${\cal H}$, there exists a densely defined closed operator $T$ in ${\cal H}$ with densely defined inverse such that $ \{ e_{n} \} \subset D(T) \cap D((T^{-1})^{\ast})$, $\phi_{n}=Te_{n}$ and $\psi_{n} =(T^{-1})^{\ast} e_{n}$, $n=0,1, \cdots$ and the minimum in such operators $T$ exists and denoted by $T_{\bm{e}}$, and furthermore there exists a unique ONB $\bm{f}= \{ f_{n} \}$ in ${\cal H}$ such that $T_{\bm{f}}$ is a non-singular positive self-adjoint operator in ${\cal H}$.\\\\
Furthermore, the author have investigated the following operators defined by a regular biorthogonal pair $( \{ \phi_{n} \} , \{ \psi_{n} \} )$ as follows: for any ONB $\bm{e}= \{ e_{n} \}$,
\begin{eqnarray}
A_{\bm{e}}
&=& T_{\bm{e}} \left( \sum_{k=0}^{\infty} \sqrt{k+1} e_{k} \otimes \bar{e}_{k+1} \right) T_{\bm{e}}^{-1} , \nonumber \\
B_{\bm{e}}
&=& T_{\bm{e}} \left( \sum_{k=0}^{\infty} \sqrt{k+1} e_{k+1} \otimes \bar{e}_{k} \right) T_{\bm{e}}^{-1}, \nonumber \\
A_{\bm{e}}^{\dagger}
&=& \left( T_{\bm{e}}^{-1}  \right)^{\ast} \left( \sum_{k=0}^{\infty} \sqrt{k+1} e_{k+1} \otimes \bar{e}_{k} \right) T_{\bm{e}}^{\ast} , \nonumber \\
B_{\bm{e}}^{\dagger}
&=& \left( T_{\bm{e}}^{-1} \right)^{\ast} \left( \sum_{k=0}^{\infty} \sqrt{k+1} e_{k} \otimes \bar{e}_{k+1} \right) T_{\bm{e}}^{\ast} ,  \nonumber \\
N_{\bm{e}}
&=& T_{\bm{e}} \left( \sum_{k=0}^{\infty} \sqrt{k+1} e_{k+1} \otimes \bar{e}_{k+1} \right) T_{\bm{e}}^{-1}, \nonumber \\
N_{\bm{e}}^{\dagger}
&=& \left( T_{\bm{e}}^{-1} \right)^{\ast}  \left( \sum_{k=0}^{\infty} \sqrt{k+1} e_{k+1} \otimes \bar{e}_{k+1} \right) T_{\bm{e}}^{\ast}, \nonumber 
\end{eqnarray}
where the tensor $x \otimes \bar{y}$ of elements $x, \; y$ of ${\cal H}$ is defined by
\begin{eqnarray}
(x \otimes \bar{y})\xi =( \xi | y)x, \;\;\; \xi \in {\cal H}. \nonumber
\end{eqnarray}
Indeed, these operators defined by ONB do not depend on methods of taking ONB. Furthermore, $A_{\bm{e}}$ and $B_{\bm{e}}$ are lowering and raising operators for $\{ \phi_{n} \}$, respectively, and $A_{\bm{e}}^{\dagger}$ and $B_{\bm{e}}^{\dagger}$ are raising and lowering operators for $\{ \psi_{n} \}$, respectively, and $N_{\bm{e}}$ and $N_{\bm{e}}^{\dagger}$ are number operators for $\{ \phi_{n} \}$ and $\{ \psi_{n} \}$, respectively, and
\begin{eqnarray}
A_{\bm{e}}B_{\bm{e}} - B_{\bm{e}}A_{\bm{e}} \subset I \;\;\;{\rm  and} \;\;\;
B_{\bm{e}}^{\dagger} A_{\bm{e}}^{\dagger} -A_{\bm{e}}^{\dagger}B_{\bm{e}}^{\dagger} \subset I . \nonumber
\end{eqnarray}
Moreover, in Ref \cite{hiroshi1} Proposition 3.4, the following statements hold with respect to the operators $A_{\bm{e}}$, $B_{\bm{e}}$, $A_{\bm{e}}^{\dagger}$ and $B_{\bm{e}}^{\dagger}$.\\
\hspace{3mm} (i) 
\begin{eqnarray}
\phi_{n}&=& \frac{1}{\sqrt{n !}} B_{\bm{e}}^{n} \phi_{0}, \;\;\; n=0,1, \cdots , \nonumber \\
\psi_{n}&=& \frac{1}{\sqrt{n!}} (A_{\bm{e}}^{\dagger})^{n} \psi_{0}, \;\;\; n=0,1, \cdots . \nonumber
\end{eqnarray}
\hspace{3mm} (ii) 
\begin{eqnarray}
A_{\bm{e}} D_{\phi}
= D_{\phi}, \;\;\; B_{\bm{e}}D_{\phi} =D_{\phi} , \nonumber
\end{eqnarray}
\hspace{9mm} and
\begin{eqnarray}
A_{\bm{e}}^{\dagger} D_{\psi}
= D_{\psi}, \;\;\;
B_{\bm{e}}^{\dagger} D_{\psi}
=D_{\psi} . \nonumber
\end{eqnarray}
Here we apply the above results to the regular biorthogonal pairs $( \{ \phi_{n} \} , \{ \psi_{n} \})$ obtained by a pseudo-bosonic operators $\{ a,b,a^{\dagger},b^{\dagger} \}$ satisfying Assumption 1-3, and may construct a new pseudo-bosonic operators $\{ A_{\bm{e}},B_{\bm{e}}, A_{\bm{e}}^{\dagger},B_{\bm{e}}^{\dagger} \}$. Therefore, we investigate the connections between the operators $a, \; b, \; a^{\dagger}, \; b^{\dagger}$ and the operators  $A_{\bm{e}}$, $B_{\bm{e}}$, $A_{\bm{e}}^{\dagger}, \; B_{\bm{e}}^{\dagger}$.

We put
\begin{eqnarray}
a_{\phi} = a \lceil _{D_{\phi}} \;\;\; &{\rm and}& \;\;\; b_{\phi}=b \lceil _{D_{\phi}}, \nonumber \\
(a^{\dagger})_{\psi} = a^{\dagger} \lceil_{D_{\psi}} \;\;\; &{\rm and}& \;\;\; (b^{\dagger})_{\psi} = b^{\dagger} \lceil_{D_{\psi}} . \nonumber
\end{eqnarray}
For relationships between $a_{\phi}, \; b_{\phi}$ and $A_{\bm{e}}, \; B_{\bm{e}}$, we have the following\\
\par
{\it Lemma 2.3.} {\it The following statements hold.\\
\hspace{3mm} (1) $a_{\phi}= A_{\bm{e}} \lceil_{D_{\phi}}$ and $b_{\phi}= B_{\bm{e}} \lceil_{D_{\phi}}$.\\
\hspace{3mm} (2) $\bar{a}_{\phi} \subset \bar{A}_{\bm{e}} $ and $\bar{b}_{\phi} \subset  \bar{B}_{\bm{e}}$.\\
\hspace{3mm} (3) If $T_{\bm{e}}^{-1}$ is bounded, then $A_{\bm{e}}$ and $B_{\bm{e}}$ are closed, that is, $\bar{a}_{\phi} \subset A_{\bm{e}} $ and $\bar{b}_{\phi} \subset B_{\bm{e}}$.\\
\hspace{3mm} (4) If $T_{\bm{e}}$ is bounded, then $\bar{a}_{\phi} = \bar{A}_{\bm{e}} $ and $\bar{b}_{\phi}= \bar{B}_{\bm{e}}$.\\
\hspace{3mm} (5) If $( \{ \phi_{n} \} , \{ \psi_{n} \} )$ is a pair of Riesz base, that is, $T_{\bm{e}}$ and $T_{\bm{e}}^{-1}$ are bounded, then $\bar{a}_{\phi} = A_{\bm{e}} $ and $\bar{b}_{\phi} = B_{\bm{e}}$.}\\
\par
{\it Proof.} (1) This follows from Proposition 2.1, (3).\\
(2) Take an arbitrary $x \in D( \bar{a}_{\phi})$. Then there exists a sequence $\{ x_{n} \}$ in $D_{\phi}$ such that $\lim_{n \rightarrow \infty} x_{n}=x$ and $\bar{a}_{\phi}x= \lim_{n \rightarrow \infty} a_{\phi} x_{n} = \lim_{n \rightarrow \infty} A_{\bm{e}} x_{n}$. Thus, we have $x \in D(\bar{A}_{\bm{e}})$ and $\bar{A}_{\bm{e}}x= \bar{a}_{\phi}x$. Therefore, we have $\bar{a}_{\phi} \subset \bar{A}_{\bm{e}}$.\\
(3) Since $A_{\bm{e}}$ and $B_{\bm{e}}$ are closed by Ref. \cite{h-t}, it follows from (1) that $\bar{a}_{\phi} \subset A_{\bm{e}} $ and $\bar{b}_{\phi} \subset B_{\bm{e}}$.\\
(4) Since $D_{\phi}$ is a core for $\bar{A}_{\bm{e}}$ and $\bar{B}_{\bm{e}}$ by Ref. \cite{hiroshi1}, it follows from (1) that $\bar{A}_{\bm{e}}= \overline{A_{\bm{e}} \lceil_{D_{\phi}} } = \bar{a}_{\phi}$ and $\bar{B}_{\bm{e}}= \overline{B_{\bm{e}} \lceil_{D_{\phi}} } = \bar{b}_{\phi}$.\\
(5) This follows from (2) and (3).\\\\
For relationships between $a_{\psi}^{\dagger}, \; b_{\psi}^{\dagger}$ and $A_{\psi}^{\dagger}, \; B_{\psi}^{\dagger}$, we have following\\
\par
{\it Lemma 2.4.} {\it The following statements hold.\\
\hspace{3mm} (1) $(a^{\dagger})_{\psi}= A^{\dagger}_{\bm{e}} \lceil_{D_{\psi}}$ and $(b^{\dagger})_{\psi}= B^{\dagger}_{\bm{e}} \lceil_{D_{\psi}}$.\\
\hspace{3mm} (2) $\bar{a}^{\dagger}_{\psi} \subset \bar{A}^{\dagger}_{\bm{e}} $ and $\bar{b}^{\dagger}_{\psi} \subset  \bar{B}^{\dagger}_{\bm{e}}$.\\
\hspace{3mm} (3) If $T_{\bm{e}}$ is bounded, then $A^{\dagger}_{\bm{e}}$ and $B^{\dagger}_{\bm{e}}$ are closed, that is,  $\bar{a}^{\dagger}_{\psi} \subset A^{\dagger}_{\bm{e}} $ and $\bar{b}^{\dagger}_{\psi} \subset B^{\dagger}_{\bm{e}}$.\\
\hspace{3mm} (4) If $T_{\bm{e}}^{-1}$ is bounded, then $\bar{a}^{\dagger}_{\psi} = \bar{A}^{\dagger}_{\bm{e}} $ and $\bar{b}^{\dagger}_{\psi}= \bar{B}^{\dagger}_{\bm{e}}$.\\
\hspace{3mm} (5) If $( \{ \phi_{n} \} , \{ \psi_{n} \} )$ is a pair of Riesz base, that is, $T_{\bm{e}}$ and $T_{\bm{e}}^{-1}$ are bounded, then $\bar{a}^{\dagger}_{\psi} = A^{\dagger}_{\bm{e}} $ and $\bar{b}^{\dagger}_{\psi} = B^{\dagger}_{\bm{e}}$.}\\
\par
{\it Proof.} The statements are proved similarly to Lemma 2.3.\\
\par
{\it Proposition 2.5.} {\it The following statements hold.\\
\hspace{3mm} (1) Suppose $D_{\phi}$ is a core for $\bar{a}$ and $\bar{b}$, then $\bar{a} \subset \bar{A}_{\bm{e}}$ and $\bar{b} \subset \bar{B}_{\bm{e}}$. In particular, if $T_{\bm{e}}^{-1}$ is bounded, then $\bar{a} \subset A_{\bm{e}}$ and $\bar{b} \subset B_{\bm{e}}$, and if $T_{\bm{e}}$ is bounded, then $\bar{a} = \bar{A}_{\bm{e}}$ and $\bar{b} = \bar{B}_{\bm{e}}$.\\
\hspace{3mm} (2) Suppose $D_{\psi}$ is a core for $\bar{a}^{\dagger}$ and $\bar{b}^{\dagger}$, then $\bar{a}^{\dagger} \subset \bar{A}^{\dagger}_{\bm{e}}$ and $\bar{b}^{\dagger} \subset \bar{B}^{\dagger}_{\bm{e}}$. In particular, if $T_{\bm{e}}$ is bounded, then $\bar{a}^{\dagger} \subset A^{\dagger}_{\bm{e}}$ and $\bar{b}^{\dagger} \subset B^{\dagger}_{\bm{e}}$, and if $T_{\bm{e}}^{-1}$ is bounded, then $\bar{a}^{\dagger} = \bar{A}^{\dagger}_{\bm{e}}$ and $\bar{b}^{\dagger} = \bar{B}_{\bm{e}}^{\dagger}$.}\\
\section{A construction of ${\cal D}$-pseudo bosons}
In Section 2, we investigate the relationships between regular biorthogonal pairs and pseudo-bosons, and the connections between the pseudo-bosonic operators $a, \; b, \; a^{\dagger}, \; b^{\dagger}$ and the pseudo-bosonic operators  $A_{\bm{e}}$, $B_{\bm{e}}$, $A_{\bm{e}}^{\dagger}, \; B_{\bm{e}}^{\dagger}$. By Proposition 2.1, (3) and Proposition 2.2, (3), $\{ a,b \}$ and $\{ a^{\dagger}, b^{\dagger} \}$ have an algebraic structure, respectively, that is, algebras generated by $\{ 1, a \lceil_{D_{\phi}}, b\lceil_{D_{\phi}} \}$ and by $\{ 1, a^{\dagger} \lceil_{D_{\psi}}, b^{\dagger} \lceil_{D_{\psi}} \}$ are defined, respectively. However, $\{ 1,a,b,a^{\dagger},b^{\dagger} \}$ do not have a $
\ast$-algebraic structure, in general. From this reason, we define the notion of ${\cal D}$-pseudo bosons in Ref \cite{bagarello13, bagarello2013} Let ${\cal D}$ be a dense subspace in a Hilbert space ${\cal H}$. We denote by ${\cal L}^{\dagger} ( {\cal D})$ the set of all linear operators $T$ from ${\cal D}$ to ${\cal D}$ such that $T^{\ast} {\cal D} \subset {\cal D}$. Here $T^{\ast}$ is the adjoint of $T$. ${\cal L}^{\dagger} ( {\cal D})$ is a $\ast$-algebra under the usual operators $\alpha T$, $S+T$, $ST$ and an involution $T^{\dagger} \equiv T^{\ast} \lceil_{{\cal D}}$, and a $\ast$-subalgebra of ${\cal L}^{\dagger} ( {\cal D})$ is called an ${\cal O}^{\ast}$-algebra on ${\cal D}$ in Ref \cite{k-s}. \\
\par
{\it Definition 3.1.} {\it A pair of operator $a$ and $b$ is ${\cal D}$-pseudo bosons if $a$, $b \in {\cal L}^{\dagger}( {\cal D})$ satisfy the following conditions (i), (ii) and (iii): \\
\hspace{3mm} (i) $ab-ba =I$. \\
\hspace{3mm} (ii) There exists a non-zero element $\phi_{0} \in {\cal D}$ such that $a\phi_{0}=0$. \\
\hspace{3mm} (iii) There exists a non-zero element $\psi_{0} \in {\cal D}$ such that $b^{\dagger} \psi_{0}=0$. }\\\\
Let ${\cal H}$ be any separable Hilbert space. We construct ${\cal D}$-pseudo bosons on the following steps:\\
\par
{\it Step 1.} Take an arbitrary ONB $\bm{e}= \{ e_{n} \}$ in ${\cal H}$. We put
\begin{eqnarray}
{\cal D} \equiv \left\{ x \in {\cal H}; \; \sum_{k=0}^{\infty} (k+1)^{n} |( x| e_{n})|^{2} < \infty, \; n=0,1, \cdots \right\} . \nonumber
\end{eqnarray}
Then ${\cal D}$ is a dense subspace in ${\cal H}$ such that
\begin{eqnarray}
\sum_{k=0}^{\infty} \sqrt{k+1} (e_{k} \otimes \bar{e}_{k+1} ) {\cal D} \subset {\cal D} \nonumber
\end{eqnarray}
and
\begin{eqnarray}
\sum_{k=0}^{\infty} \sqrt{k+1} (e_{k+1} \otimes \bar{e}_{k} ) {\cal D} \subset {\cal D}. \nonumber
\end{eqnarray}
\par
{\it Step 2.} Take an arbitrary operator $T$ and the inverse $T^{-1}$ in ${\cal L}^{\dagger} ( {\cal D})$.\\
\par
{\it Step 3.} We define operators $A$ and $B$ on ${\cal D}$ by
\begin{eqnarray}
A
&=& T \left( \sum_{k=0}^{\infty} \sqrt{k+1} e_{k} \otimes \bar{e}_{k+1}  \right) T^{-1}, \nonumber \\
B
&=& T \left( \sum_{k=0}^{\infty} \sqrt{k+1} e_{k+1} \otimes \bar{e}_{k}  \right) T^{-1}. \nonumber
\end{eqnarray}
Then,
\begin{eqnarray}
A^{\dagger}
&=& \left( T^{-1} \right)^{\dagger} \left( \sum_{k=0}^{\infty} \sqrt{k+1} e_{k+1} \otimes \bar{e}_{k}  \right) T^{\dagger}, \nonumber \\
B^{\dagger}
&=& \left( T^{-1} \right)^{\dagger} \left( \sum_{k=0}^{\infty} \sqrt{k+1} e_{k} \otimes \bar{e}_{k+1}  \right) T^{\dagger}, \nonumber
\end{eqnarray}
and 
\begin{eqnarray}
AB-BA=I \;\;\; {\rm on} \; {\cal D}. \nonumber
\end{eqnarray}
Furthermore, we put
\begin{eqnarray}
\phi_{n}=Te_{n} , \;\;\; \psi_{n}=(T^{-1})^{\dagger} e_{n}, \;\;\; n=0,1, \cdots . \nonumber
\end{eqnarray}
Then $( \{ \phi_{n} \} , \{ \psi_{n} \} )$ is a regular biorthogonal pair, and $A$ and $B$ are lowering and raising operators for $\{ \phi_{n} \}$, respectively, and $A^{\dagger}$ and $B^{\dagger}$ are raising and lowering operators for $\{ \psi_{n} \}$, respectively. Thus we have the following\\
\par
{\it Theorem 3.2.} {\it For any ONB $\bm{e}= \{ e_{n} \}$ in ${\cal H}$, we may construct operators $A$ and $B$ satisfying ${\cal D}$-pseudo bosons.}\\\\
By a method of taking an ONB $\bm{e}= \{ e_{n} \}$ and a operator $T$ satisfying Step 2, we can construct many different ${\cal D}$-pseudo bosons.
\section{Examples of ${\cal D}$-pseudo bosons}
In this section, we give some physical examples of ${\cal D}$-pseudo bosons constructed by Step 1-3 in Section 3. We may construct various models by a method of taking the ONB $\bm{f}= \{ f_{n} \}$ in the Hilbert space $L^{2}( \bm{R})$ consisting of 
\begin{eqnarray}
f_{n}(x)
= \frac{1}{\sqrt{2^{n} n!} } H_{n}(x) e^{- \frac{1}{2} x^{2}}, \nonumber
\end{eqnarray}
where $H_{n}(x)$ is the $n$th Hermite polynomial, and an operator $T$ in $L^{2}(\bm{R})$ satisfying Step 2.\\

Let $S(\bm{R})$ be the Schwartz space of all infinitely differentiable rapidly decreasing functions on $\bm{R}$. The ONB $\bm{f}= \{ f_{n} \}$ is contained in $S(\bm{R})$. We define the momentum operator $p$ and the position operator $q$ by
\begin{eqnarray}
&&\left\{
\begin{array}{cl}
& D(p): \; {\rm the \; set \; of \; all \;  differentiable \; functions} \; f \; {\rm on} \; \bm{R} \; {\rm such \; that} \; \frac{df}{dx} \in L^{2}(\bm{R}), \\
\nonumber \\
& (pf)(x) = -i \frac{df}{dx}, \;\;\; f \in D(p),
\end{array}
\right. \nonumber \\
\nonumber \\
&&\left\{
\begin{array}{cl}
& D(q)= \left\{ f \in L^{2}(\bm{R}); \; \int_{-\infty}^{\infty} |xf(x)|^{2} dx < \infty \right\}, \\
\nonumber \\
& (qf)(x) = xf(x), \;\;\; f \in D(q).
\end{array}
\right. \nonumber
\end{eqnarray}
These operators $p$ and $q$ are self-adjoint operators in $L^{2}(\bm{R})$ and $S(\bm{R})$ is a core for $p$ and $q$, and furthermore $p S(\bm{R}) \subset S(\bm{R})$ and $q S(\bm{R}) \subset S(\bm{R})$. We introduce the standard bosonic operators $a= \frac{1}{\sqrt{2}} (q+ i p)$ and $a^{\dagger}= \frac{1}{\sqrt{2}} (q- i p)$, which obey $[a,a^{\dagger}]=I$. Here we consider some examples\\
\par
{\it Example 1.} (The extended quantum harmonic oscillator)\\
Let $\beta >0$. We put
\begin{eqnarray}
T= e^{- \frac{1}{\beta^{2}}} e^{\frac{a+a^{\dagger}}{\beta}}=  e^{-\frac{1}{\beta^{2}}\frac{\sqrt{2}}{\beta}q}. \nonumber
\end{eqnarray}
Then $T$ and $T^{-1}$ are self-adjoint operators in $L^{2}(\bm{R})$ such that $TS(\bm{R}) \subset S(\bm{R})$ and 
$T^{-1} S(\bm{R}) \subset S(\bm{R})$. We denote the restriction $T$ to $S(\bm{R})$ by the same $T$. Thus the ONB $\bm{f}= \{ f_{n} \}$ in $L^{2}(\bm{R})$ and $T$ satisfy the conditions of Step 1-3 in Section 3 for ${\cal D} \equiv S(\bm{R})$. Hence we may construct ${\cal D}$-pseudo bosonic operators
\begin{eqnarray}
A
&=& T \left( \sum_{k=0}^{\infty} \sqrt{k+1} f_{k} \otimes \bar{f}_{k+1} \right) T^{-1}, \nonumber \\
B
&=& T \left( \sum_{k=0}^{\infty} \sqrt{k+1} f_{k+1} \otimes \bar{f}_{k} \right) T^{-1}. \nonumber 
\end{eqnarray}
Then it is shown that $A=a- \frac{1}{\beta}$ and $B=a^{\dagger}+ \frac{1}{\beta}$, and the non self-adjoint hamiltonian $H_{\beta} = \frac{\beta}{2} (p^{2}+q^{2}) + \sqrt{2} i p$, introduced in Ref \cite{bagarello2013} and Ref \cite{bagarello13}, can be written by
\begin{eqnarray}
H_{\beta}
= \beta \left( BA+ \frac{2+ \beta^{2} }{2 \beta^{2}} I \right). \nonumber
\end{eqnarray}\\
\par
{\it Example 2.} (The Swanson model)\\
Let $\theta$ be a real parameter value in $(- \frac{\pi}{4}, \frac{\pi}{4}) \setminus \{ 0 \}$. We put
\begin{eqnarray}
T= e^{i \frac{\theta}{2} (a^{2}-a^{\dagger 2}) }  =e^{- \frac{\theta}{2} (q p +p q ) }. \nonumber
\end{eqnarray}
Then $T$ and $T^{-1}$ are self-adjoint operators in $L^{2}(\bm{R})$ such that $TS(\bm{R}) \subset S(\bm{R})$ and 
$T^{-1} S(\bm{R}) \subset S(\bm{R})$. We denote the restriction $T$ to $S(\bm{R})$ by the same $T$. Thus the ONB $\bm{f}= \{ f_{n} \}$ in $L^{2}(\bm{R})$ and $T$ satisfy the conditions of Step 1-3 in Section 3 for ${\cal D} \equiv S(\bm{R})$. Hence we may construct ${\cal D}$-pseudo bosonic operators
\begin{eqnarray}
A
&=& T \left( \sum_{k=0}^{\infty} \sqrt{k+1} f_{k} \otimes \bar{f}_{k+1} \right) T^{-1}, \nonumber \\
B
&=& T \left( \sum_{k=0}^{\infty} \sqrt{k+1} f_{k+1} \otimes \bar{f}_{k} \right) T^{-1}. \nonumber 
\end{eqnarray}
Then it is shown that $A=\cos( \theta) \; a+ i \sin (\theta) \; a^{\dagger}$ and $B=\cos (\theta) \; a^{\dagger} + i \sin (\theta) \; a$, and the non self-adjoint hamiltonian $H_{\theta} = \frac{1}{2} (p^{2}+q^{2}) -  \frac{i}{2} \tan (2 \theta) (p^{2}-q^{2})$, introduced in Ref \cite{bagarello13} and Ref \cite{j-n-j}, can be written by
\begin{eqnarray}
H_{\theta}
= \frac{1}{\cos(2 \theta)} \left( BA+ \frac{1 }{2 } I \right). \nonumber
\end{eqnarray}\\
In next Example 3 and 4 we shall take the operators $T$ that are different to those in Example 1 and 2, respectively to construct various models.\\
\par
{\it Example 3.} Let $\beta >0$. We put
\begin{eqnarray}
T= e^{ \frac{1}{\beta^{2}}} e^{i \frac{a-a^{\dagger}}{\beta}}=  e^{-\frac{1}{\beta^{2}}\frac{\sqrt{2}}{\beta}p}. \nonumber
\end{eqnarray}
Then $T$ and $T^{-1}$ are self-adjoint operators in $L^{2}(\bm{R})$ such that $TS(\bm{R}) \subset S(\bm{R})$ and 
$T^{-1} S(\bm{R}) \subset S(\bm{R})$. We denote the restriction $T$ to $S(\bm{R})$ by the same $T$. Thus the ONB $\bm{f}= \{ f_{n} \}$ in $L^{2}(\bm{R})$ and $T$ satisfy the conditions of Step 1-3 in Section 3 for ${\cal D} \equiv S(\bm{R})$. Hence we may construct ${\cal D}$-pseudo bosonic operators
\begin{eqnarray}
A
&=& T \left( \sum_{k=0}^{\infty} \sqrt{k+1} f_{k} \otimes \bar{f}_{k+1} \right) T^{-1}, \nonumber \\
B
&=& T \left( \sum_{k=0}^{\infty} \sqrt{k+1} f_{k+1} \otimes \bar{f}_{k} \right) T^{-1}. \nonumber 
\end{eqnarray}
Then it is shown that $A=a+ \frac{i}{\beta}$ and $B=a^{\dagger}+ \frac{i}{\beta}$.\\
\par
{\it Example 4.} Let $\theta$ be a real parameter value in $(- \frac{\pi}{4}, \frac{\pi}{4}) \setminus \{ 0 \}$. We put
\begin{eqnarray}
T= e^{ \frac{\theta}{2} (a^{2}-a^{\dagger 2}) }   =e^{i \frac{\theta}{2} (q p +p q ) }. \nonumber
\end{eqnarray}
Then $T$ and $T^{-1}$ are non self-adjoint operators in $L^{2}(\bm{R})$ such that $TS(\bm{R}) \subset S(\bm{R})$ and 
$T^{-1} S(\bm{R}) \subset S(\bm{R})$. We denote the restriction $T$ to $S(\bm{R})$ by the same $T$. Thus the ONB $\bm{f}= \{ f_{n} \}$ in $L^{2}(\bm{R})$ and $T$ satisfy the conditions of Step 1-3 in Section 3 for ${\cal D} \equiv S(\bm{R})$. Hence we may construct ${\cal D}$-pseudo bosonic operators
\begin{eqnarray}
A
&=& T \left( \sum_{k=0}^{\infty} \sqrt{k+1} f_{k} \otimes \bar{f}_{k+1} \right) T^{-1}, \nonumber \\
B
&=& T \left( \sum_{k=0}^{\infty} \sqrt{k+1} f_{k+1} \otimes \bar{f}_{k} \right) T^{-1}. \nonumber 
\end{eqnarray}
Then it is shown that $A=\cosh ( \theta) \; a+  \sinh (\theta) \; a^{\dagger}$ and $B=\cosh(\theta) \; a^{\dagger} +  \sinh (\theta) \; a$.

\ \\
Graduate School of Mathematics, Kyushu University, 744 Motooka, Nishi-ku, Fukuoka 819-0395, Japan
\\
h-inoue@math.kyushu-u.ac.jp, 

\ \\
Department of Applied Mathematics, Fukuoka University, Fukuoka 814-0180, Japan\\
mayumi@fukuoka-u.ac.jp, \\


\begin{thebibliography}{99}
\bibitem{h-t}
H. Inoue and M. Takakura,
Non-self-adjoint hamiltonians defined by generalized Riesz bases,
e-print., arXiv:math-ph/1604.00161

\bibitem{hiroshi1}
H. Inoue,
General theory of regular biorthogonal pairs and its physical applications,
e-print., arXiv:math-ph/1604.01967

\bibitem{k-s}
K. Schm{\"u}dgen,
Unbounded Operator Algebras and Representation Theory,
Birkh{\"a}user-Verlag., Basel, 1990

\bibitem{b-i-t}
F. Bagarello,
Non-self-adjoint hamiltonians defined by Riesz bases,
J. Math. Phys., {\bf 55}(2014), 033501

\bibitem{bagarello2013}
F. Bagarello and A. Inoue and C. Trapani,
From self to non self-adjoint harmonic oscillators: physical consequences and mathematical pitfalls,
Phys. Rev. A., {\bf 88}(2013), 032120


\bibitem{bagarello13}
F. Bagarello,
More mathematics for pseudo-bosons,
J. Math. Phys., {\bf 54}(2013), 063512

\bibitem{bagarello11}
 F. Bagarello, (Regular) pseudo-bosons versus bosons, J. Phys. A., {\bf 44}(2011), 015205

\bibitem{bagarello10}
F. Bagarello,
Pseudobosons, Riesz bases, and coherent states,
J. Math. Phys., {\bf 51}(2010), 023531


\bibitem{mostafazadeh}
 A. Mostafazadeh, 
 Pseudo-Hermitian representation of Quantum Mechanics, Int. J. Geom. Methods Mod. Phys., {\bf 7}(2010), 1191-1306 

\bibitem{j-n-j}
 J. da Provid{\^e}ncia and N. Bebiano and J. P. da Provid{\^e}ncia, 
 Non hermitian operators with real spectrum in quantum mechanics, ELA., {\bf 21}(2010), 98-109 

\bibitem{d-t}
D.A. Trifonov,
Pseudo-boson coherent and Fock states,
e-print., arXiv:quant-ph/0902.3744

\end{thebibliography}
\end{document}